\documentclass[reprint,superscriptaddress,aps,pra,twocolumn,floatfix]{revtex4-2}
\usepackage[T1]{fontenc}
\usepackage[latin1]{inputenc}
\usepackage{lmodern}
\expandafter\let\csname equation*\endcsname\relax
\expandafter\let\csname endequation*\endcsname\relax
\usepackage{amsmath}
\usepackage{amssymb}
\usepackage{graphicx}
\usepackage{xcolor}
\usepackage{natbib}
\usepackage{bm}
\usepackage{dsfont}
\usepackage{bbold}
\usepackage[mathscr]{euscript}
\usepackage[cal=boondoxo]{mathalfa}
\usepackage{comment}
\renewcommand\footnotemark{}

\usepackage{braket}
\usepackage[normalem]{ulem}
\usepackage{soul}

\def\ii{{\rm i}}

\def\ge{\sigma_{ge}}  
  
\def\hge{\hat{\sigma}_{ge}}  
\def\heg{\hat{\sigma}_{eg}}

\def\ga{\Gamma_{\rm 1D}}
\def\kk{k_{\rm 1D}}
\def\gap{\Gamma'}

\def\bra#1{\mathinner{\langle{#1}|}}
\def\ket#1{\mathinner{|{#1}\rangle}}
\def\braket#1{\mathinner{\langle{#1}\rangle}}
\def\oj{\tilde{\omega}_0}
\def\Dj{\tilde{\Delta}}
\def\dej{\tilde{\delta}_\xi}

\def\zl{z_\text{left}}  
\def\zr{z_\text{right}} 

\begin{document}
\title{Optical precursors in waveguide quantum electrodynamics}
\author{S. Cardenas-Lopez}
\affiliation{Department of Physics, Columbia University, New York, NY 10027, USA}
\author{P. Solano}
\affiliation{Departamento de F\'isica, Facultad de Ciencias F\'isicas y Matem\'aticas, Universidad de Concepci\'on, Concepci\'on, Chile}
\author{L. A. Orozco}
\affiliation{Joint Quantum Institute, Department of Physics and NIST, University of Maryland, College Park, MD 20742, USA}
\author{A. Asenjo-Garcia}
\email{ana.asenjo@columbia.edu}
\affiliation{Department of Physics, Columbia University, New York, NY 10027, USA}
\date{\today}

\begin{abstract}
When a broadband signal propagates through a dispersive medium, some frequency components move faster than the center of the pulse. This leads to the appearance of precursors, transient signals that emerge from the medium earlier than the main part of the pulse and seem to propagate superluminally. Here, we investigate the microscopic origin of precursors in a minimal setup: an array of qubits coupled to a waveguide. The linear transmission function only converges to that of a continuous medium for large qubit numbers. Nevertheless, the dispersion produced by only two qubits is enough to produce oscillatory transients. Precursors are best observed under conditions of electromagnetically-induced transparency, as the center of the pulse is significantly delayed. Under these conditions, just a single qutrit is enough to generate a precursor. Our results pave the way towards dispersion engineering of light with just a few qubits, and can be realized with superconducting qubits coupled to transmission lines or atoms coupled to optical waveguides.
\end{abstract}
\maketitle

\section{Introduction}
The propagation of light through a continuous dispersive medium is a canonical problem in electrodynamics \cite{Burnham69,Crisp70}. One of the most fascinating aspects of the transmitted radiation is the formation of transients that precede and follow the main pulse when the input signal has sharp edges, compared to the response time of the system. These transients are known as precursors and were theoretically predicted by Sommerfeld and Brillouin in the early 20th century~\cite{Sommerfeld, Brillouin}. They occur due to the different group velocities of the high- and low-frequency components in the spectrum of the original pulse. Since their prediction, precursors have been extensively studied theoretically~\cite{PhysRevA.104.013518,Macke12,Macke09,Macke13}. They have also been observed in a plethora of systems and frequency ranges, such as in microwaves propagating in transmission lines~\cite{PhysRevLett.22.1201}, optical photons traversing atomic clouds~\cite{EIT_flash,PhysRevA.56.1564,PhysRevLett.96.143901}, gamma rays in M{\"o}ssbauer spectroscopy~\cite{PhysRev.124.1178, int,Adams_gammarays,rohlsberger_gammarays,PhysRevLett.66.770} and mechanical waves in fluids \cite{Avenel83,Falcon03}. 

In waveguide quantum electrodynamics (wQED), where a collection of qubits are coupled to each other via a one-dimensional (1D) photonic reservoir, dispersion arises due to the narrow spectral response of the qubits. These two-level systems (which represent neutral atoms, molecules, or superconducting qubits, to name a few examples) only interact with photons of frequency resonant with their ground- to excited-state transition. Photon-mediated interactions between qubits give rise to the emergence of collective states that can either decay rapidly (superradiant) or be protected from dissipation (subradiant)~\cite{Albrecht_2019}. In the last decade, wQED has attracted significant interest due to the possibility of exploiting these states for quantum information processing and storage (for instance, to produce quantum states of light~\cite{GonzalezTudela15PRL} and to compute via decoherence-free subspaces~\cite{Paulisch_2016}) as well as for exploring many-body physics in open quantum systems (many-body localization~\cite{PhysRevResearch.3.033233}, spin dimerization~\cite{chiral}, and fermionization~\cite{Albrecht_2019}, among other examples). A lot of work has been devoted to single-~\cite{Chang07}, few-~\cite{Shen07,Roy11,Shi11,XuS13,Fang15,Ke19}, and many-photon~\cite{PhysRevX.10.031011} transport in wQED, although most of it (except for a few exceptions~\cite{Liao15,buchpaper,pennetta2021observation}) is focused on the steady-state regime or on propagation of quasi-monochromatic light. In parallel, experimental realizations of wQED systems have multiplied, with setups ranging from neutral atoms coupled to single-mode optical fibers~\cite{Vetsch10, Goban12, Gouraud15, Solano17} and photonic-crystal waveguides~\cite{Thompson13,Goban15,Hood16}, to superconducting qubits coupled to microwave transmission lines~\cite{Liu16,Mirhosseini19}. 

Here, we investigate the transport of broadband photon pulses in wQED, for a system consisting of $N$ qubits coupled to a waveguide. Employing a transmission coefficient in terms of collective frequency shifts and decay rates, we demonstrate that the temporal response under a short pulse coincides with that of a continuous medium for $N\gg 1$. The macroscopic description of the medium breaks down for a small qubit number. Nevertheless, just two qubits generate enough dispersion to produce an intensity profile that oscillates rapidly in time. The delay between the main signal and its precursor is evident under conditions of electromagnetically induced transparency, where a single qutrit is enough to generate a precursor. 



\section{Continuous medium}
The amplitude and phase modulation acquired by a monochromatic field after propagation through a (either classical or quantum) linear dispersive medium is encoded in the complex transmission function $t(\omega)$. For non-monochromatic input pulses, the transmitted field is simply 
\begin{equation}
E(t)=\frac{1}{2\pi}\int_{-\infty}^\infty  t(\omega) E_0(\omega) e^{-\ii\omega t}d\omega,
\label{eq:presc}
\end{equation}
where $E_0(\omega)$ is the Fourier transform of the temporal profile of the input pulse. In standard electromagnetism, the complex relative permittivity $\epsilon(\omega)$ determines the transmission coefficient, and is usually postulated phenomenologically to match the optical response of a continuous medium. A conventional model is that of a Lorentz oscillator with a single resonance of frequency $\omega_0$ and damping coefficient $\Gamma'\ll \omega_0$. Then, the transmission coefficient reads \cite{stratton}
\begin{equation}\label{lorentz}
t_\text{cont}(\omega)=\text{exp}\left(-\frac{\ii b}{\omega-\omega_0+\ii\Gamma'/2}\right),
\end{equation}
where $b$ quantifies the strength of the light-matter interaction. Throughout this paper, we consider an input field of central frequency $\omega_p$ with a square temporal profile, $E_0(t)=\mathcal{E}_0e^{\ii\omega_p t}\left[\Theta(t-t_\text{i})-\Theta(t-t_\text{f})\right]$. Assuming that the duration of the input pulse is larger than the time it takes the system to reach the steady state, we approximate the input signal  as a step function  $E_0(t)\simeq \mathcal{E}_0 e^{-i\omega_p t}\Theta(t_f-t)$ to calculate the transients in the transmitted intensity right after switching off the input field. Similarly, to calculate the transmitted field immediately after switching on the input field, we  approximate the pulse as $E_0(t)\simeq \mathcal{E}_0e^{-i\omega_p t}\Theta(t-t_i)$. Setting $t_f=0$ ($t_i=0$), the Fourier transform for the rising (falling) edge is

\begin{equation}
E_{0, \{R,F\}}(\omega)=\mathcal{E}_0\bigg[\pm \mathcal{P}\frac{\ii}{(\omega-\omega_p)}+\pi\delta(\omega-\omega_p)\bigg],
\label{eq:fourier}
\end{equation}

where $+(-)$ corresponds to the raising (falling) edge and $\mathcal{P}$ stands for Cauchy principal value. Plugging $E_{0, \{R,F\}}(\omega)$ and $t_\text{cont}(\omega)$ into the transmitted field expression in Eq.~\eqref{eq:presc} results in

\begin{eqnarray}\label{eq:integral}
E_{\{R,F\}}(t)&&=\mathcal{E}_0\big[\Theta(t_f-t)e^{-\ii\omega_p t}t_\text{cont}(\omega_p)\\\nonumber
&&\mp\tfrac{1}{2\pi i} e^{-\ii\omega_0 t}e^{-\Gamma't/2}\oint \frac{dz}{z-\Delta-\ii\Gamma'/2}e^{-\ii \frac{b}{z}}e^{-\ii z t}\big],
\end{eqnarray}
where $z\equiv\omega-\omega_0+\ii\Gamma'/2$ and $\Delta=\omega_p-\omega_0$ is the detuning between the central and resonance frequencies. After solving the integral (see Appendix \ref{appendix1}), the final form for the transients reads

\begin{eqnarray}
\label{eq:bessels}
	&&\frac{I_\text{cont}(t)}{I_0}=\bigg|\Theta(t_f-t)t_\text{cont}(\omega_p)e^{-\ii\Delta(t-t_0)}-e^{-\Gamma' (t-t_0)/2}\nonumber\\ 
	&&\times\sum_{n=1}^\infty \left(\frac{-\ii}{\Delta+\ii\Gamma'/2}\sqrt{\frac{b}{(t-t_0)}}\right)^nJ_n\left(2\sqrt{ b (t-t_0)}\right)\bigg|^2,
\end{eqnarray}

where $J_n$ is a Bessel function of the first kind. Here, $t_0=t_i$ corresponds to the rising edge and $t_0=t_f$ to the falling one. The sharp edges of the input signal translate into a broad spectrum in Fourier space, and the interference between different frequency components propagating at different group velocities gives rise to temporal oscillations in the transmitted intensity. The transmitted intensity consists only of the precursor and the final transient, since the main pulse has been absorbed and scattered to free space.

\section{$N$ Qubits}

\subsection{Model}

\begin{figure}
\centering
\includegraphics[width=0.5
\textwidth]{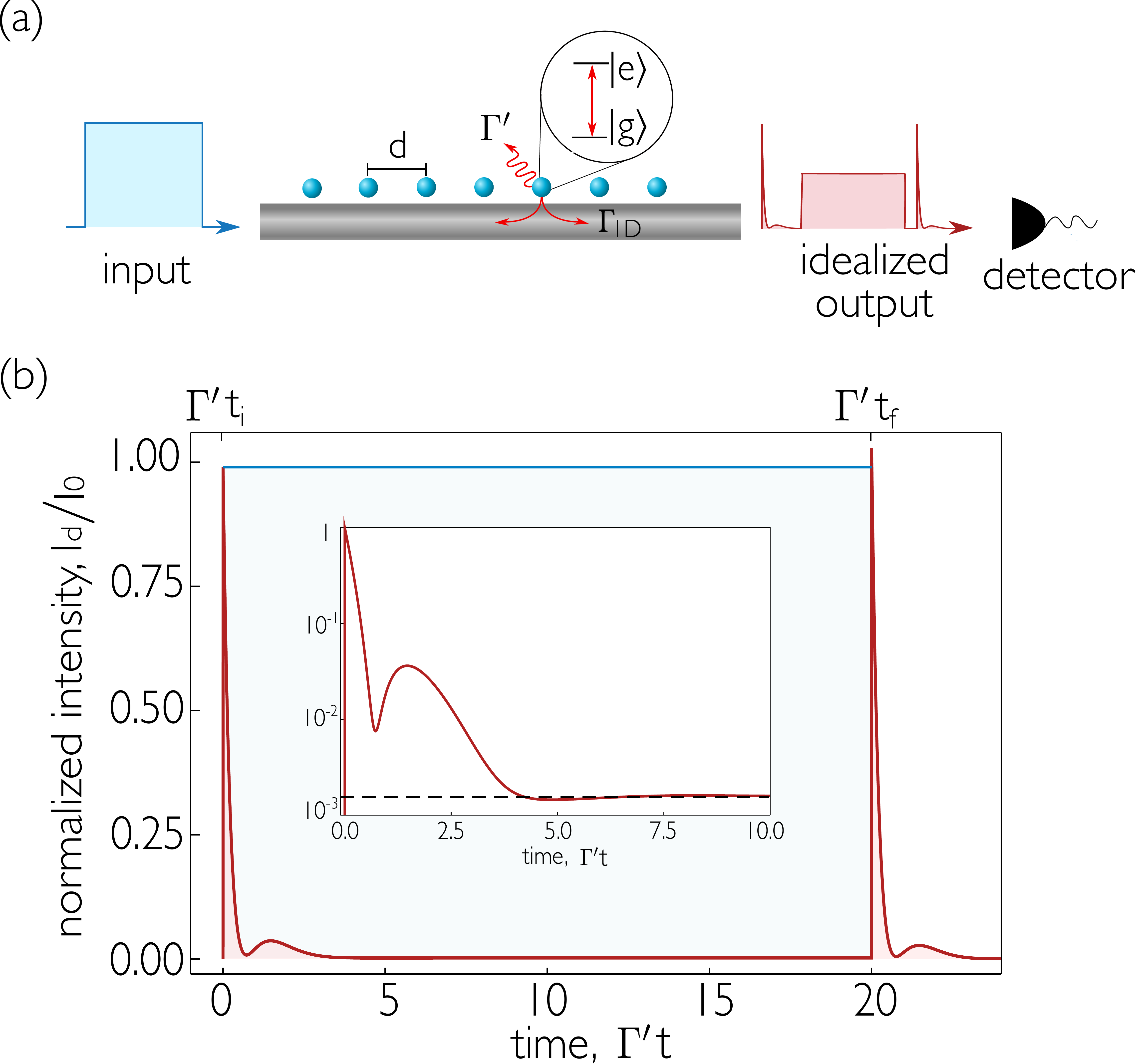}
\caption{Propagation of a broadband signal in a waveguide-QED system. (a) A sharp input pulse propagates through an array of qubits separated by a distance $d$. The qubit decay rate into the waveguide is $\ga$, and the decay rate into any other mode is $\Gamma'$. The idealized output consists of a precursor followed by the main pulse and a trailing transient.  (b) Temporal evolution of the intensity at the detector, $I_d$ with (red) and without (blue) qubits, normalized to the maximum input intensity $I_0$. The central part of the pulse is mostly absorbed, and the transmitted intensity is low except at the edges, where a precursor can be observed at the first edge. The qubits' response immediately after switch-on and off is transient, taking some time for the induced field to build up (and cancel the forward-propagating input field) and decay, respectively.  The optical depth is $N\Gamma_\text{1D}/\Gamma'=5$ ($N=20$, $\Gamma_\text{1D}/\Gamma'=0.25$), the detuning between the resonance and central frequency of the pulse is $\Delta=0.37\Gamma'$, the lattice constant is $\kk d=\pi/2$, and $t_{i(f)}$ represents the time at which the pulse is switched on (off).}
\label{fig:fig1}
\end{figure}

We demonstrate that temporal oscillations in the transmitted field intensity (so-called ``dynamical beats'' in the M{\"o}ssbauer literature~\cite{PhysRev.124.1178, int,Adams_gammarays,rohlsberger_gammarays,PhysRevLett.66.770}) are not unique to continuous classical media, but also occur in ``granular'' quantum systems, such as a chain of $N>1$ qubits coupled to a 1D waveguide, as shown in Fig.~\ref{fig:fig1}. In this case, the optical response can be obtained by first tracing out the field and solving for the dynamics of the qubits~\cite{spinmodel}, which interact with each other as they share a common electromagnetic environment. Then, one recovers the field evolution via an input-output formalism. In the linear (or single excitation) regime, the qubits' evolution is governed by the effective Hamiltonian  $\mathcal{H}=\mathcal{H}_\text{1D}+\mathcal{H}'+\mathcal{H}_\text{drive}$ \cite{Chang_12,Lalumiere13,Caneva15}, where
\begin{subequations}
\begin{gather}
\mathcal{H}_\text{1D}=-\ii \frac{\hbar\ga}{2}\sum_{i,j=1}^N e^{\ii\kk d |i-j|} \heg^i\hge^j, \\
\mathcal{H}'=\hbar\left(J'-\ii\frac{\gap}{2}\right)\sum_{i=1}^N\hat{\sigma}^i_{ee},\\
\mathcal{H}_\text{drive}=-\hbar\Delta\sum_{i=1}^N\hat{\sigma}^i_{ee}-\hbar\Omega(t)\sum_{i=1}^N \left(e^{\ii \kk z_i}\heg^i +\text{H. c.}\right).
\end{gather}
\label{eq:ham}
\end{subequations}

Here, $\mathcal{H}_\text{1D}$ describes the qubit-qubit interaction, which occurs at a rate $\ga$, depends on the lattice constant $d$, and is mediated by photons of wave-vector $\kk$. The qubits may decay (independently from each other) into other non-guided modes at a rate $\Gamma'$, and the presence of the waveguide imparts a Lamb shift $J'$ on their resonance frequency, as described by $\mathcal{H}'$. The qubit ensemble is being driven by a propagating pulse of Rabi frequency $\Omega(t)$, whose central frequency is detuned from the qubit resonance frequency by $\Delta=\omega_p-\omega_0$. In the above equations, $\heg^i=\ket{e_i}\bra{g_i}$ is the coherence operator between the $i$th-qubit excited and ground states, $\hat{\sigma}_{ee}^i=\ket{e_i}\bra{e_i}$, and H.c. stands for Hermitian conjugate. We note that the rotating wave approximation is justified as counter-rotating terms produce rapidly oscillating contributions (at frequency $\sim 2\omega_0$) that average out in the timescales relevant for the transients (proportional to the inverse of the qubit linewidth). The transmitted light intensity is found via the input-output equation \cite{Lalumiere13,Caneva15}

\begin{equation}
   \hat{E}^+(z, t)=\Omega(t) e^{\ii\kk z}+\ii\frac{\ga}{2}\sum_{i=1}^N e^{\ii\kk |z-z_i|}\hge^i(t),
\end{equation}

where $\hat{E}^+$ is the positive-frequency component of the right-propagating field (normalized to have units of Rabi frequency), and the field is measured by a detector at a position $z$ that lies beyond the last qubit. In this manuscript, the dispersion is solely due to the qubits, and the waveguide is considered to be  dispersionless.

\subsection{Transmission coefficient}
\begin{figure*}
\centering
\includegraphics[width=1
\textwidth]{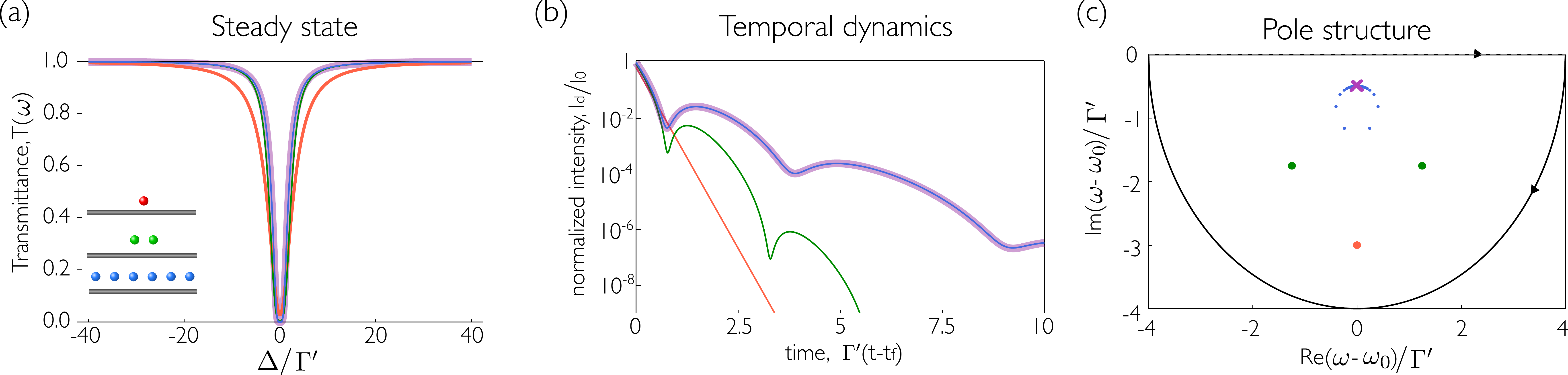}
\caption{Steady-state and time-dependent optical response for 1 (red), 2 (green), 200 (blue) qubits and for a continuous medium (purple). (a) Transmittance as a function of frequency for a continuous-wave input. (b) Light intensity right after switching-off a square pulse at $t=t_f$, as a function of time. (c) The temporal response can be understood by analyzing the pole structure of the transmission coefficient in the complex plane. The response function for a finite qubit array, $t_N$, has simple poles at finite real frequencies for $N\geq 2$. As $N\rightarrow\infty$, the poles (blue dots) converge towards the essential singularity of the continuous transmission coefficient $t_\text{cont}$ at $\omega_*=\omega_0-\ii\gap/2$, indicated by a purple cross. For all plots, $\kk d=\pi/2$ and $\Delta=0.3\gap$. The coupling efficiency $\ga/\gap$ is changed for each qubit number, keeping the optical depth the same ($OD\equiv N\ga/\gap=5$).}
\label{fig:fig2}
\end{figure*}

The steady-state transmission is mostly determined by the optical depth $OD\equiv N\ga/\Gamma'$, as shown in Fig.~\ref{fig:fig2}(a), where systems with different number of qubits and decay rates but fixed optical depth display almost identical transmittance spectra. To calculate the transmitted light for a continuous wave drive (i.e., $E_0(t)=\Omega_0e^{\ii\omega_p t}$), we solve for the expectation value of the steady-state coherences (such that $\braket{\dot{\hat{\sigma}}_{eg}^i}=0$) and plug the result into the above input-output equation. The transmission coefficient is defined as 
\begin{equation}
    t_N(\omega)=\frac{E^+(\zr)}{E_p^+(\zl)},
\end{equation}
where $\zl$ is a point immediately to the left of the qubit~1 and $\zr$ is a point immediately to the right of the qubit~$N$. $E^+(z)$ is the expectation value of the positive-frequency component of the total field operator $\hat{E}(z)$ in the steady state and $E_p^+(z)=\Omega_0 e^{\ii\kk z}$ is the input field. The transmission coefficient can be expressed in terms of collective shifts and decay rates, as we now derive (see ~\cite{PRAAna} for full details).

The input-output equation states that the total field is the sum of the input field and the field radiated by the qubits, i.e.,
\begin{equation}
E^+(z)=\Omega_0 e^{\ii\kk z}-\sum_{n=1}^N g(z,z_n)\ge^n,
\end{equation}
where $\ge^n\equiv \langle\hat{\sigma}_{ge}^n\rangle$ is the expectation value for the coherences in the steady state and $g(z,z')=-\ii(\ga/2) e^{\ii\kk|z-z'|}$. Defining $\mathfrak{g}_{nm}=g(z_n,z_m)$, and  $\textbf{E}^+_{p}$ a $N$-dimensional vector whose entries are $E^+_{p,n}\equiv E_p^+(z_n)$, the evolution equations for the expectation value of the coherences  are
\begin{equation}
\dot{\sigma}_{ge}^n=\ii\bigg(\Delta-J'+\ii\frac{\gap}{2}\bigg)\ge^n+i E^+_{p,n}-\ii\sum_{m=1}^N\mathfrak{g}_{nm}\ge^m.
\end{equation}

The steady state solutions of these equations (for which $\dot{\sigma}_{ge}^n=0$) are
\begin{equation}
    \Vec{\sigma}_{ge}=-\mathcal{M}^{-1}\textbf{E}^+_{p},
    \label{eq:st}
\end{equation}
with  $\mathcal{M}=\left(\Delta-J'+\ii\frac{\gap}{2}\right)\mathbb{1}-\mathfrak{g}$. We express this in terms of collective modes, since  the eigenvectors of $\mathfrak{g}$ satisfy $\sum_{\xi=1}^N\textbf{v}_\xi\otimes \textbf{v}_\xi^T=\mathds{1}$. Using this identity, we find
\begin{equation}
    \Vec{\sigma}_{ge}=-\sum_{\xi=1}^N \frac{\left(\textbf{v}_\xi^T\cdot \textbf{E}^+_{p}\right)\textbf{v}_\xi}{\Delta-J'+\ii\gap/2-\lambda_\xi},
    \label{eq:st2}
\end{equation}

where $\{\lambda_\xi\}$ are the eigenvalues of $\mathfrak{g}$ (and of $\mathcal{H}_\text{1D}$ defined in Eq.~\eqref{eq:ham}). Plugging the steady state solution ~\eqref{eq:st2}  into the expression for the field we obtain 
\begin{equation}
E^+(z)=E_p^+(z)+\sum_{\xi=1}^N\frac{(\textbf{g}(z)\cdot \textbf{v}_\xi)(\textbf{v}_\xi^T\cdot \textbf{E}^+_{p})}{\Delta-J'+\ii\gap/2-\lambda_\xi}.
\end{equation}

In the last expression we have adopted the shorthand notation $(\textbf{g}(z))_j=g(z,z_j)$. Here, $g(z,z')$ is the propagator of the guided field that has been projected in the direction of the qubits' dipole transition. Physically, $g(z,z')$ describes the field at $z$ that is generated by a dipole at $z'$. By means of the trace-determinant lemma~\cite{PRAAna}, this expression can be written in terms of eigenvalues only, yielding

\begin{equation}
   t_N(\omega)= \prod_{\xi=1}^N \frac{\omega-\oj+\ii\Gamma'/2}{\omega-\oj+\ii\Gamma'/2-\lambda_\xi}.
   \label{eq:response_func}
\end{equation}

where $\oj=\omega_0+J'$. The real ($J_\xi=\text{Re}\{\lambda_\xi\}$) and imaginary ($\Gamma_\xi=-2\text{Im}\{\lambda_\xi\}$) parts of these eigenvalues correspond to the frequency shifts and decay rates of the collective modes. The decay rates can be either superradiant (with $\Gamma_\xi>\ga$, and the largest one scaling as $\Gamma_\xi\sim N\ga$) or subradiant (with $\Gamma_\xi<\ga$, and the smallest one scaling as $\Gamma_\xi\sim \ga/N^3$), and their actual values depend on the specific lattice constant \cite{Albrecht_2019,PhysRevResearch.3.033233}. For lattice constants such that $\kk d=n\pi$, with $n$ being an integer, there is only one superradiant eigenvalue of decay $N\ga$. In this so-called `mirror configuration', there is only one collective mode coupled to the waveguide, and the array of qubits behaves effectively as a single qubit with a large decay rate \cite{PRAAna}.

\subsection{\label{disc_transients} Transients}

The temporal dynamics of the transmitted intensity for a broadband input pulse is not determined solely by the optical depth, but is instead sensitive to the specific values of $N$ and $\ga/\Gamma'$ separately, as shown in Fig.~\ref{fig:fig2}(b). From Eqs.~\eqref{eq:presc},~\eqref{eq:fourier} and~\eqref{eq:response_func}, the transmitted field at the beginning and end of the pulse is

\begin{eqnarray}
E_{\{R, F\}}(t)&&=\mathcal{E}_0\big[\frac{1}{2}e^{-\ii\omega_p t}t_N(\omega_p)\\\nonumber
&&\mp\frac{1}{2\pi \ii} \mathcal{P}\int_{-\infty}^{\infty}d\omega \frac{1}{\omega-\omega_p}t_N(\omega)e^{-\ii\omega t} \big].
\label{eq:tfield1}
\end{eqnarray}

We solve the remaining integral using the residue theorem and a semicircle that closes in the lower-half plane. There are two contributions: one from the simple pole at $\omega=\omega_p$, and one due to the singularities in $t(\omega)$ in the lower-half plane. They read

\begin{eqnarray}
    I_1&&=-\pi \ii \,\text{Res}\left(\frac{1}{\omega-\omega_p}t_N(\omega)e^{-\ii\omega t},\omega_p\right)\\\nonumber
    &&=-\pi \ii\, t_N(\omega_p)e^{-\ii\omega_p t},
\end{eqnarray}
\begin{equation}
    I_2=-2 \pi \ii \sum_{\xi}\text{Res}\left(\frac{1}{\omega-\omega_p}t_N(\omega)e^{-\ii\omega t},\omega_\xi\right),
\end{equation}
where $\{\omega_\xi\}$ are the singularities of $t_N(\omega)$. 

Reintroducing the times $t_i$ and $t_f$  yields the final expression for the transmitted field 



\begin{eqnarray}
\label{eq:unified}
	&&E_{\{R, F\}}(t)=\mathcal{E}_0\big[\Theta(t_f-t)e^{-i\omega_p (t-t_0)}t_N(\omega_p)\nonumber\\ 
	&&\pm\sum_{\xi}\text{Res}\left(\frac{1}{\omega-\omega_p}t_N(\omega)e^{-\ii\omega (t-t_0)},\omega_\xi\right)\big],
\end{eqnarray}

\begin{figure}
\centering
\includegraphics[width=0.45\textwidth]{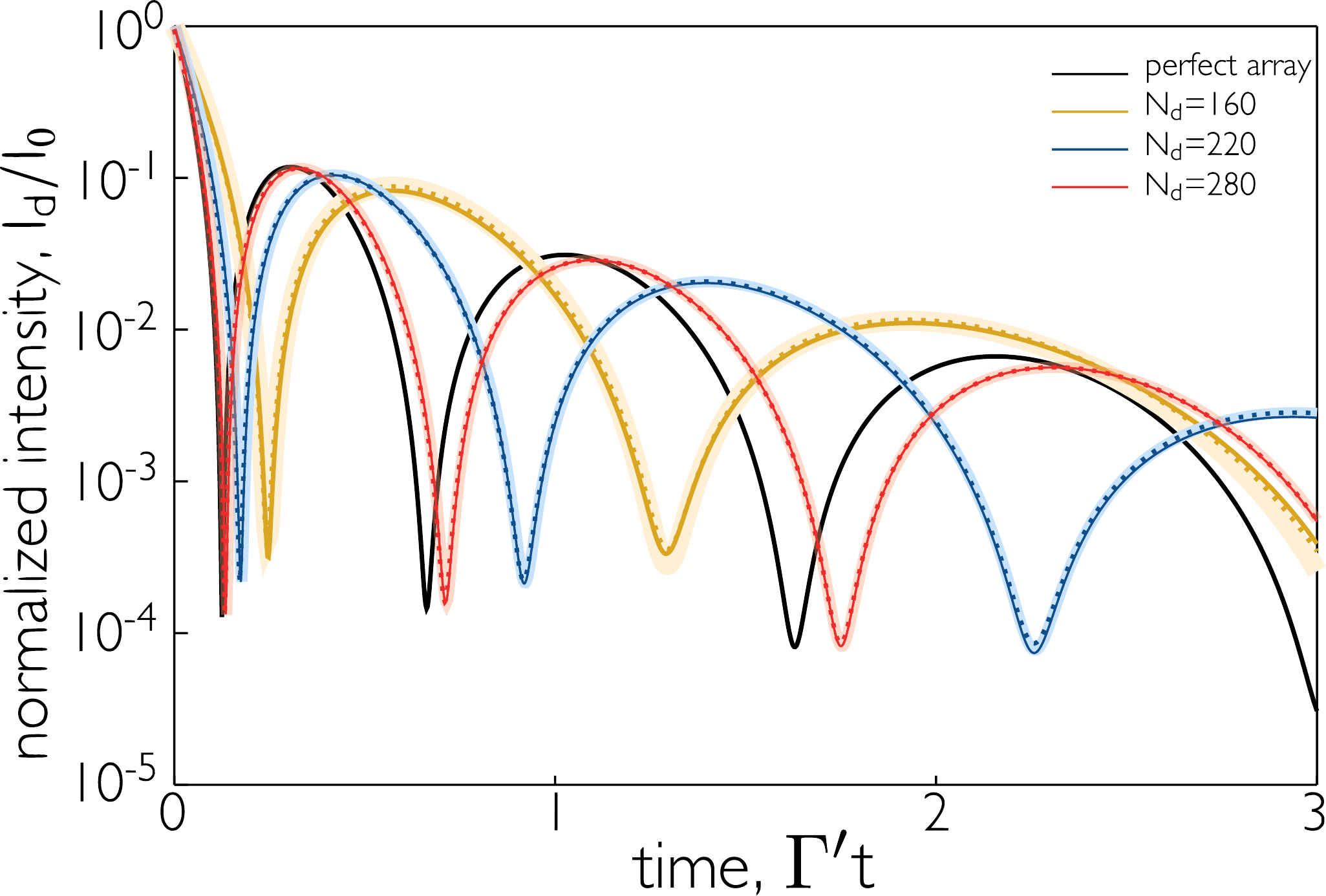}
\caption{Light intensity after switching off a square pulse propagating through an imperfectly-filled array with 300 sites, $\ga/\Gamma'=0.08$ and $\Delta=0.3\Gamma'$. In each run, there are $N_d$ qubits, with $N_d\leq N$. The results of the imperfectly-filled arrays (solid lines), perfect arrays of $N_d$ qubits (thick light lines), and the Bessel expansion of Eq.~\eqref{eq:bessels} with $b=N_d\ga/2$ (dotted lines) are in agreement.}
\label{fig:figdis}
\end{figure}

where $t_0=t_i$ at the rising edge and $t_0=t_f$ at the falling edge.

The term originated by the simple pole $\xi$ in the transmission coefficient for a discrete chain, $t_N(\omega)$, describes the light emitted by the corresponding  collective mode. The measured intensity at the rising and falling edges is thus a coherent sum over all the contributions of the collective modes and reads

\begin{eqnarray}
    \frac{I_N (t)}{I_0}&&=\bigg|\Theta(t_f-t)t_N(\omega_p)e^{-i\Dj t}+e^{-\Gamma' (t-t_0)/2}\nonumber\\
    &&\times\sum_{\xi=1}^N \frac{\lambda_\xi^N}{\prod_{\kappa\neq \xi}(\lambda_\xi-\lambda_\kappa)}\frac{e^{-i\lambda_\xi (t-t_0)}}{\lambda_\xi-\Dj-\ii\Gamma'/2}\bigg|^2,
    \label{eq:finiteatoms}
\end{eqnarray}
where $\Dj=\omega_p-\oj$. The contributions from different modes give rise to a time-dependent slope in the decay. The most superradiant modes play an important role for shorter times. As these modes become depopulated, the most significant contributions originate from less superradiant states, yielding a smaller decay rate ~\cite{SVIDZINSKY20092894}. The transients cannot be faithfully reproduced by just including a few modes in Eq.~\eqref{eq:finiteatoms}.

Oscillations only occur for $N\geq 2$, as they arise from interference between different collective modes. Two qubits is the minimum required number to produce oscillations that -- only in this case -- are periodic with a frequency equal to half of the difference between the two collective modes. For $N\gg1$, the intensity calculations agree with those obtained for a continuous medium [i.e., as described by Eq.~\eqref{eq:bessels}]. The agreement can be readily understood by noting that for $|\lambda_\xi|\ll|\omega-\omega_0+\ii \gap/2|$,  with $\xi=\{1,...,N\}$, the transmission coefficient for a finite array reduces to 
\begin{equation}
t_{N\gg 1}(\omega)=e^{-\sum_\xi\ln [1-\lambda_\xi/(\omega-\tilde{\omega}_0+\ii\Gamma'/2)]}\simeq e^{{\sum\limits_{\xi}}\frac{\lambda_\xi}{\omega-\oj+\ii\Gamma'/2}},
    \label{eq:mac_response_func}
\end{equation}
which is precisely the transmission coefficient of a continuous Lorentz medium (Beer-Lambert law), as captured by Eq.~\eqref{lorentz}, with resonant frequency $\omega_0+J'$, damping coefficient $\Gamma'$, and coupling strength $b=|\sum_{\xi} \lambda_\xi|=N\ga/2$. For a fixed optical depth, the region where the series expansion is valid increases with qubit number [see Fig. ~\ref{fig:fig2}(c)], thus the approximation works better for $N\gg 1$. As exemplified in Fig. \ref{fig:figdis}, in this limit the temporal evolution of the intensity is independent of the qubit spatial configuration, is robust against imperfect filling of the array, and is dictated only by the optical depth. This occurs for any lattice constant different from that of the mirror configuration.

Moreover, Eq.~\eqref{eq:bessels} captures the temporal response even for large optical depths (as long as $N\gg 1$, $N\ga/\Gamma'$ can take any value), as shown in Fig.~\ref{fig:fig2}(b) for $N=200$ qubits and $N\ga/\Gamma'=5$, even if $t_N$ cannot be approximated by $t_\text{cont}$ at resonance. This occurs because the temporal response for a broadband pulse involves an integral over frequencies, which is less sensitive to the specific details of the response function at resonance, compared to the steady-state transmission.

The breakdown of the continuous approximation for a few qubits can be understood by analyzing the pole structure of the two transmission coefficients in the complex plane, as shown in Fig.~\ref{fig:fig2}(c). For a finite array, each qubit (or more specifically, each collective mode) contributes with one simple pole. As $N \rightarrow \infty$, the poles cluster around $\omega_*=\tilde{\omega}_0-\ii \Gamma'/2$, which is an essential singularity of $t_\text{cont}(\omega)$, and the response coincides to that of a continuous medium. As the number of qubits decreases, the poles do not densely cover the region of the essential singularity. Nevertheless, due to the frequency splitting between the collective modes, the poles have finite real parts, producing oscillations in the time domain. Lastly, $N=1$ has a single pole at a purely imaginary frequency, $-\ii(\ga+\Gamma')/2$, giving rise to exponential, non-oscillatory decay. The approximation in Eq.~\eqref{eq:mac_response_func} is still valid for a single atom as long as $\ga\ll\gap$. However, in this limit, $t_{N\gg 1}(\omega)$ roughly predicts a purely exponential decay, since the oscillations would occur in a timescale ($\sim \ga^{-1}$) that is much larger than the decay ($\sim \gap^{-1}$).

Eq.~\eqref{eq:finiteatoms} describes the transients only when the Markov approximation is valid, i.e., when retardation is negligible, and when the bandwidth of the reservoir is much larger than the linewidth of the qubits [thus making $\ga$ and $k_\text{1D}$ approximately constants in the frequency interval with the most important contributions to the integral in  Eq.~\eqref{eq:presc}]. If this approximation is not valid [for instance, for total chain lengths comparable to or larger than $c/(\gap+\ga)$], precursors are expected to appear even at the mirror configuration. The output field in this regime is

\begin{figure}
\centering
\includegraphics[width=0.4\textwidth]{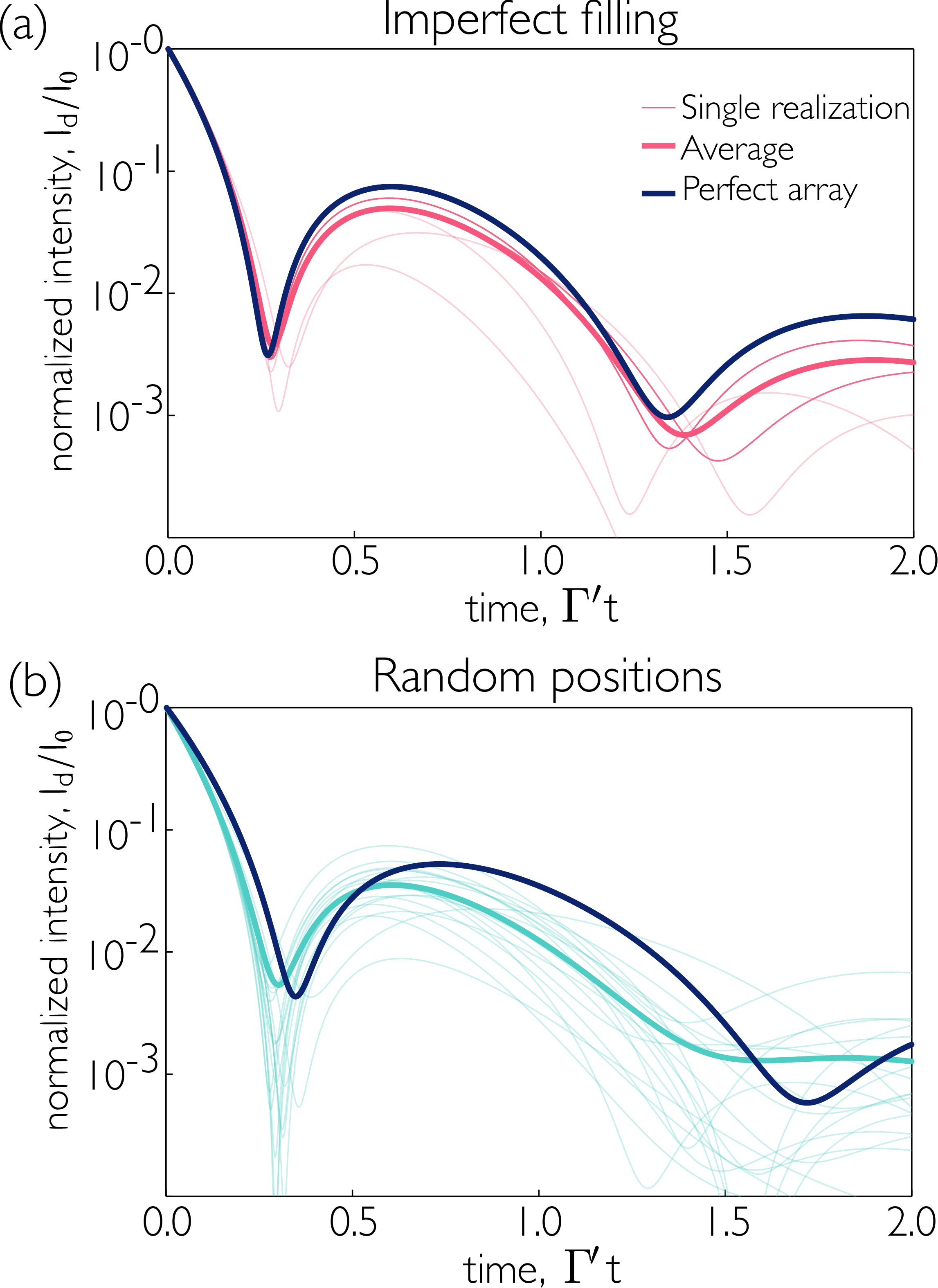}
\caption{Impact of position disorder on transients for  a low qubit number. (a) Transients produced by 5 atoms positioned at random in an array of 10 sites with lattice constant $\kk d=\pi/2$ . (b) Transients produced by 5 atoms with random positions and $\kk z_\text{max}=5\pi$. $\ga/\gap=2.5$ and  $\Delta/\gap=0.5$ for both plots. 100 realizations were used to get the average curves.}
\label{fig:figdis2}
\end{figure}

\begin{equation}
E(t)=\frac{1}{2\pi}\int_{-\infty}^\infty  \tilde{t}(\omega) E_0(\omega) e^{-\ii\omega (t-z_\text{right}/c)}d\omega,
\end{equation}
where $\tilde{t}(\omega)$ is the transmission coefficient in the non-Markovian limit. The transmission coefficient takes the form~\cite{PRAAna},
\begin{equation}
\tilde{t}(\omega)= 1- \tfrac{1}{g(z_\text{right},z_\text{left},\omega)}\sum_{\xi}\tfrac{\left(\textbf{g}(z_\text{right},\omega)\cdot\textbf{v}_\xi(\omega)\right)\left(\textbf{v}_\xi^T(\omega)\cdot \textbf{g}(z_\text{left},\omega)\right)}{\tilde{\Delta}+\ii \gap/2-\lambda_\xi(\omega)}.
\label{eq:t2}
\end{equation}

Here,  $g(z,z',\omega)=-[\ii\ga(\omega)/2] e^{\ii\kk(\omega)|z-z'|}$ with  $(\textbf{g}(z,\omega))_j=g(z,z_j,\omega)$. In the non-Markovian regime, both the eigenvalues $\{\lambda_\xi(\omega)\}$ and eigenvectors $\{\textbf{v}_\xi(\omega)\}$ of $g_{ij}=g(z_i,z_j,\omega)$ depend on frequency. Even if the qubits are in the mirror configuration at resonance (i.e., $\kk(\omega_0)d=n\pi$), $\tilde{t}(\omega)$ will generically have more than one simple pole, since $\kk(\omega)d\neq n\pi$ for most frequencies, so $g_{ij}$ will have more than one nonzero eigenvalue. Hence the contributions from the singularities of $\tilde{t}(\omega)$ to the output field will interfere and produce oscillations.

\subsection{Role of spatial disorder and inhomogeneous broadening}

The transients for high qubit numbers depend exclusively on the optical depth. For low qubit numbers, the transmitted intensity is not solely dictated by the OD. As shown in Fig. \ref{fig:figdis2},  transients change for different spatial configurations. The transients produced by the perfect array can be approximately recovered by averaging over many realizations with imperfect arrays with the same qubit number, as shown in Fig.~\ref{fig:figdis2}(a). Even if the qubits have random positions, as in Fig.~\ref{fig:figdis2}(b), the average over many realizations is qualitatively similar to the signal of a perfect array.

\begin{figure}\centering
\includegraphics[width=0.4\textwidth]{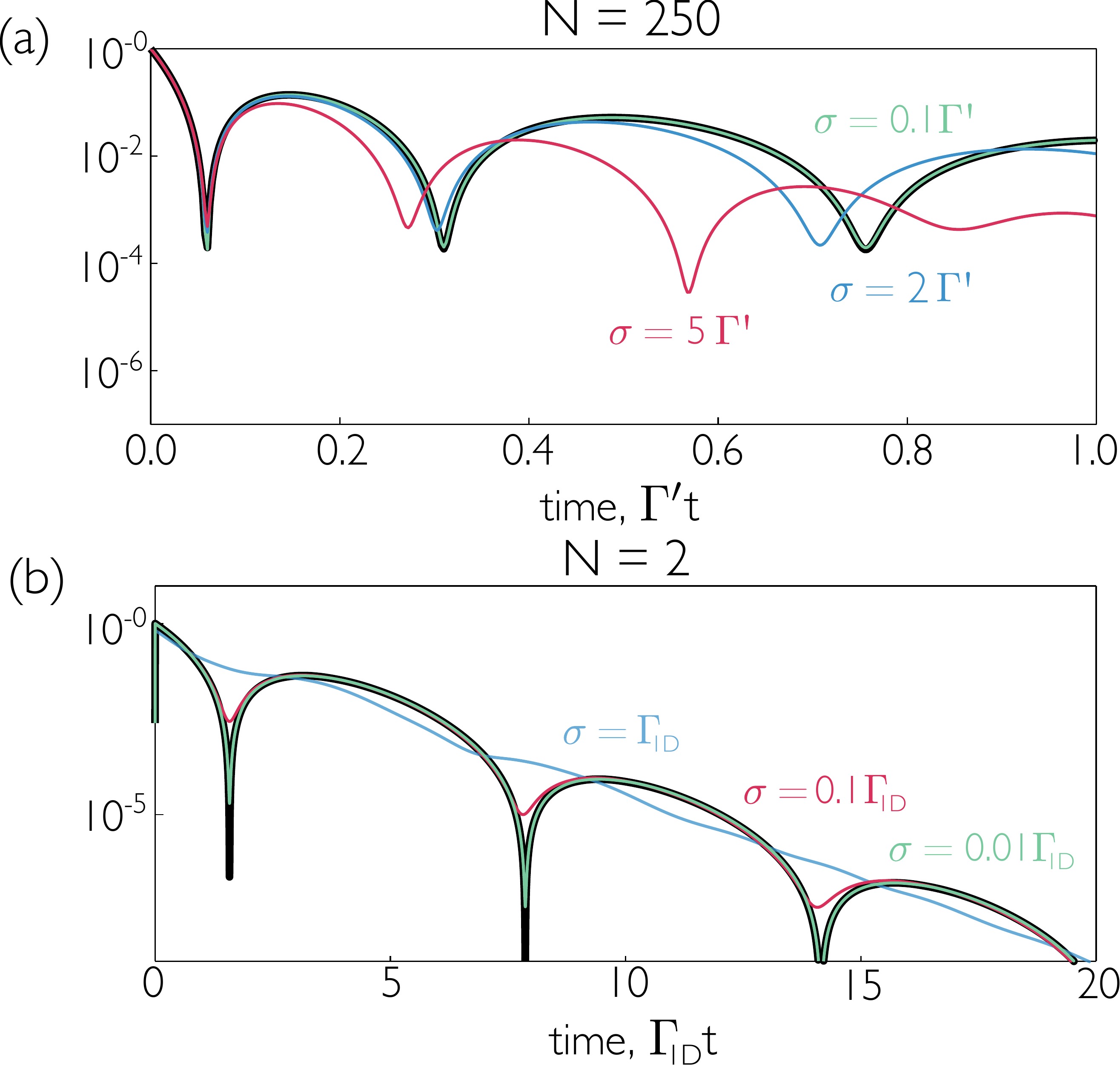}
\caption{Impact of frequency disorder on transients for (a) large and (b) small qubit numbers. The parameters used were respectively ($\ga/\gap=0.2$, $\Delta/\gap=0$, $\kk d=\pi/2$)  and  ($\ga/\gap=25$, $\Delta/\gap=0$, $\kk d=\pi/2$). The curves shown for every $\sigma$ are the result of averaging over 100 realizations.}
\label{fig:figdis3}
\end{figure}

Next, we analyze the effect of inhomogeneous broadening on the  transients. We consider that atom $i$ is detuned from the central frequency of the input pulse by $\Delta_i$, which is chosen randomly from a Gaussian distribution of mean $\mu=0$ and standard deviation $\sigma$. As shown in Fig. \ref{fig:figdis3}, the transients described by Eqs.~\eqref{eq:bessels} and 
~\eqref{eq:finiteatoms} are robust against typical disorder levels found in experimental realizations for both large ( $\sigma \sim \gap$ \cite{Hood16}) and low qubit numbers ($\sigma \sim 0.01\ga$ \cite{Mirhosseini19}).

\section{$N$ Qutrits}

\begin{figure*}\centering
\includegraphics[width=\textwidth]{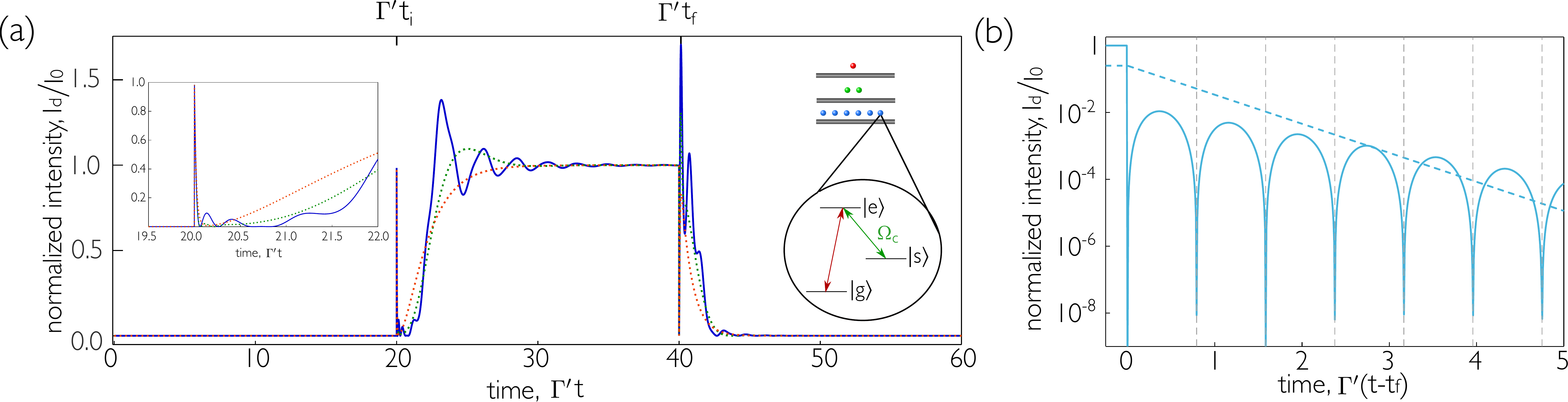}
\caption{Electromagnetically-induced transparency enables clear observation of the full-time evolution of the precursors as it delays the main pulse. (a) Transmitted intensity for 1 (red), 2 (green), and 50 (blue) qutrits with $\kk d= \pi/2$, where EIT is achieved by coupling an external control field to the transition between the excited and metastable $\ket{s}$ states, as shown in the schematics. The inset shows the precursors emerging at early times. The optical depth is $N\ga/\gap=50$.  (b) Oscillations in intensity right after the pulse has been switched off at $t=t_f$, for a single qutrit. The gray dashed lines show the oscillation period (see main text for details). For comparison,  the decay for $\Omega_c=0$ is shown as a dashed blue line. The optical depth is $N\ga/\gap=1$.  In both plots, the calculations are done in the condition of two-photon resonance, i.e., $\Delta_s=\tilde{\Delta}$, with $\tilde{\Delta}=0$. The control field intensity is $\Omega_c=4\Gamma'$.}
\label{fig:fig3}
\end{figure*}

 To observe the delay between the main pulse and its precursor, one can employ qutrits (three-level systems) under electromagnetic-induced transparency (EIT) conditions~\cite{Macke09}. EIT has been used to observe precursors in both coherent~\cite{EIT_flash} and single-photon~\cite{zhang} pulses propagating through dilute atomic clouds in free space. By coupling the excited state to a metastable level $\ket{s}$ via a control field of Rabi frequency $\Omega_c$, a transparency window of width $\sim \Omega_c^2/\sqrt{N\ga\Gamma'}$ opens up and a pulse that is spectrally narrower than the window propagates without being absorbed or reflected,  at a reduced group velocity $v_g=2\Omega_c^2 d/\ga$. As shown in Fig.~\ref{fig:fig3}(a), a square pulse is also delayed and the precursor is measured before the main signal arrives at the detector. The system is described by the Hamiltonian
\begin{equation}
\mathcal{H}_{\text{EIT}}=\mathcal{H}_\text{1D}+\mathcal{H}'-\hbar \sum_{n=1}^N \left(\Delta_s \hat{\sigma}_{ss}^i+\Omega_c \left(\hat{\sigma}_{es}^i+\hat{\sigma}_{se}^i\right)\right),
\end{equation}
where $\Delta_s=\omega_p-\omega_c-\omega_s$ is the two-photon detuning (with $\omega_c$ being the frequency of the control field) and $\hat{\sigma}^i_{es}=\ket{e_i}\bra{s_i}$ is the coherence operator between the excited and the metastable states.

The transmission coefficient describing light propagation through this system is derived in an analogous manner to $t_{N}(\omega)$  and reads~\cite{PRAAna}
\begin{equation}
t_{\text{EIT}}(\omega)=\prod_{\xi=1}^N\frac{(\omega-\oj)(\omega-\oj+\ii\Gamma'/2)-\Omega_c^2}{(\omega-\oj)(\omega-\oj+\ii\Gamma'/2-\lambda_\xi)-\Omega_c^2}.
\label{eq:teit}
\end{equation}

 This response function has $2N$ poles at (complex) frequencies
\begin{equation}
    \omega_\xi^\pm=\oj+\frac{\dej}{2}\pm\sqrt{\dej^2/4+\Omega_c^2},
\end{equation}

where $\dej\equiv \lambda_\xi-\ii\gap/2$. Proceeding in a similar way as in section~\ref{disc_transients}, the transmitted intensity is
\begin{eqnarray}
    \frac{I_d}{I_0}&&=\bigg|\Theta(t_f-t)t_\text{EIT}(\omega_p)e^{-\ii\omega_p(t-t_0)}\\ \nonumber
    &&+\sum_{\nu=1}^N \text{Res}\bigg(\frac{t_\text{EIT}(\omega)e^{-i\omega (t-t_0)}}{\omega-\omega_p},\omega_\nu^+\bigg)\\ \nonumber
    &&+\text{Res}\bigg(\frac{t_\text{EIT}(\omega)e^{-i\omega (t-t_0)}}{\omega-\omega_p},\omega_\nu^-\bigg)\bigg|^2.
\end{eqnarray}

Plugging in the poles and simplifying results in
\begin{widetext}
\begin{eqnarray}
    \frac{I_d(t)}{I_0}&&=\bigg|\Theta(t_f-t)t_\text{EIT}(\omega_p)e^{-\ii\tilde{\Delta}(t-t_0)}+2e^{-\Gamma'(t-t_0)/4}\sum_{\xi=1}^N\frac{\lambda_\xi^{N}}{\Omega_\xi\prod_{\eta\neq\xi}(\lambda_\xi-\lambda_\eta)}\frac{e^{-\ii\lambda_\xi (t-t_0)/2}}{\Omega_\xi^2-(\lambda_\xi-\ii\Gamma'/2-2
    \Dj)^2}\nonumber\\
    &&\times\left[2\Omega_\xi \tilde{\Delta}\cos \frac{\Omega_\xi (t-t_0
    )}{2}-2\ii\left(\tilde{\Delta}(\lambda_\xi-\ii\Gamma'/2)+2\Omega_c^2\right)\sin\frac{\Omega_\xi (t-t_0)}{2}\right]\bigg|^2,
    \label{eq:eit_tpulse}
\end{eqnarray}
\end{widetext}

with  $\Omega_\xi=\sqrt{4\Omega_c^2+\tilde{\delta}_\xi^2}$. The transmitted field after a large number of qutrits consists of an initial precursor, the main pulse, and a final transient. For large enough optical depth, the precursor is clearly separated from the (delayed) main pulse, as shown in Fig.~\ref{fig:fig3}(a).

At resonance, the intensity of the precursor is always the intensity of the main pulse  ($I_d(t=t_i)=I_0$), as can be seen from Eq.~\eqref{eq:eit_tpulse}. Furthermore, the delay time of the main pulse can be inferred by estimating the timescale of decay of the second term. For simplicity, we consider $\tilde{\Delta}=0$. The dependence in time of the contribution of mode $\nu$ to the second term of Eq.~\eqref{eq:eit_tpulse} is

\begin{eqnarray} \label{eq:modenu}
    E_\nu(t) &&\sim e^ {-\gap (t-t_i)/4} e^ {-\ii (t-t_i)\lambda_\nu/2}\sin\Omega_\nu (t-t_i)/2\nonumber\\ 
    &&\sim e^ {-\gap (t-t_i)/4} e^ {-\ii (t-t_i)\lambda_\nu/2}\nonumber\\ 
    &&\times\left(e^{\ii\Omega_\nu (t-t_i)/2}-e^{-\ii\Omega_\nu (t-t_i)/2}\right).
\end{eqnarray}

The most important contributions to Eq.~\eqref{eq:eit_tpulse} come from the most superradiant modes, for which we can neglect the shift and approximate  $\lambda_\nu=J_\nu -\ii\Gamma_\nu/2\sim -\ii\Gamma_\nu/2$.  Furthermore, we can assume $\Gamma_\nu\gg \Gamma',\Omega_c$, which yields
$$
\ii\text{Im}\Omega_\nu=\ii\text{Im}\sqrt{4\Omega_c^2+(\lambda_\nu-\ii \Gamma'/2)^2}\simeq \frac{\ii \Gamma_\nu}{2}-\frac{4\ii\Omega_c^2 }{\Gamma_\nu}.
$$
Plugging this expression into Eq.~\eqref{eq:modenu}, we find

\begin{equation}
E_\nu(t)\sim  e^{-\Gamma_\nu t/2}e^{2\frac{\Omega_c^2 }{\Gamma_\nu}t}-e^{-2\frac{\Omega_c^2 }{\Gamma_\nu}t}\simeq -e^{-2\frac{\Omega_c^2 t}{\Gamma_\nu}},
\end{equation}
where the first term has been neglected as $\Gamma_\nu\gg \Omega_c$. The term with the slowest decay thus belongs to the most superradiant mode, for which $\Gamma_\nu \simeq N\ga$. Hence, the time of arrival of the main pulse (i.e., the time at which the output signal stabilizes to $I_0$) scales as $\sim \frac{ N\ga}{\Omega_c^2}$, which agrees with the delay of a monochromatic wave propagating through a dilute atomic cloud~\cite{RevModPhys.77.633,EIT_flash}.

Under conditions of EIT, a single qutrit is enough to produce both a precursor and oscillations in the intensity after switching-off the input field, as shown in Fig.~\ref{fig:fig3}(b). Due to the coupling of the excited and metastable state, each qutrit contributes to two poles (of finite and opposite frequency) to the transmission coefficient. Right after switching off the field, there is no radiation as the excited state is unpopulated. The emission of light follows a slower trend, as light can only be emitted when the qutrits oscillate into the excited state. In the limit where $\Omega_c\gg \ga$, the oscillations have a period that scales as $\sim 1/\sqrt{4\Omega_c^2-(\gap+\ga)^2/4}$. The dips in the intensity correspond to times when the population of $\ket{e}$ is minimal. For more qutrits, the oscillations are almost periodic if $\Omega_c\gg N\ga$, or resemble those of two-level systems (Fig.~\ref{fig:fig2}) in the opposite case.

\section{Conclusions}

In summary, waveguide QED constitutes a versatile platform for dispersion engineering, which can be employed to tailor the temporal shape of propagating photons. The transport of spectrally-broad photon pulses can be understood from the location of the poles of transmission coefficient in the complex plane, which correspond to the collective frequency shifts and decay rates arising from photon-mediated qubit-qubit interactions. For large qubit number, the response of the discrete system approaches that of a continuous medium, where the temporal oscillations in the intensity (arising from interference between frequency components of the original pulse) are fully determined by the optical depth $N\ga/\Gamma'$. In contrast, for low qubit number, there is a breakdown of the macroscopic response, and a single qutrit is enough to give rise to precursors (under EIT conditions). In this regime, the optical depth is no longer a good figure of merit, and the Purcell factor and the number of qubits, separately, play a significant role in the dynamics.

Precursors provide information about the number of qubits and their coupling separately, in contrast to the Beer-Lambert law that is obeyed by systems under continuous-wave illumination. For low atom numbers, this feature allows one to count how many atoms are coupled to a nanostructure, which is difficult to do with continuous-wave measurements \cite{Hood16}. Both the oscillation timescale for the transients and the delay between precursor and main pulse in the EIT regime increase with the ratio $\ga/\gap$. Efficient coupling to the waveguide mode can be achieved in state-of-the-art experimental platforms, such as in superconducting qubits coupled to microwave transmission lines~\cite{Mirhosseini19} ($\ga/\gap>100$) and quantum dots coupled to photonic crystal waveguides ($\ga/\gap\simeq 10$~\cite{Lund-Hansen08,Arcari14}).

\textbf{Acknowledgments} -- The authors acknowledge D. H. Phong, B. Macke, and S. Du for useful discussions. We gratefully acknowledge support from the Air Force Office of Scientific Research through their Young Investigator Prize (grant No.~21RT0751), the A. P. Sloan foundation, and the David and Lucile Packard foundation. S. C.-L. acknowledges support from the Chien-Shiung Wu Family Foundation. This work was supported in part by CONICYT-PAI grant 77190033, FONDECYT grant N$^{\circ}$ 11200192 from Chile.

\bibliography{apssamp}

\appendix

\section{\label{appendix1}Contour integral for the transients in a continuous medium}	

We follow closely the calculation presented in Ref.~\cite{int} to solve the integral in Eq.~\eqref{eq:integral}. Since  $\sum\limits_{k=0}^\infty r^k=1/(1-r)$, we replace the denominator by the series
\begin{equation}
\frac{1}{z-\Delta-\ii\Gamma'/2}=-\sum_{k=0}^\infty\frac{z^k}{(\Delta+\ii\Gamma'/2)^{k+1}}.
\end{equation}
We also replace the exponentials inside the contour integral by the generating function of the Bessel functions of the first kind:
\begin{equation}
e^{-\ii b/z}e^{-\ii zt}=\sum_{m=-\infty}^{\infty}\left(-\ii z\sqrt{\frac{t}{b}}\right)^m J_m(2\sqrt{tb}).
\end{equation}

The field at the trailing edge is therefore

\begin{eqnarray}
	&&E_F(t)=-\frac{\mathcal{E}_0}{2\pi i} e^{-\ii\omega_0 t}e^{-\Gamma't/2}\sum_{m=-\infty}^{\infty}\sum_{k=0}^\infty\oint \frac{dz \,z^{k+m}}{(\Delta+\ii\Gamma'/2)^{k+1}}\nonumber\\ 
	&&\times\left(-\ii\sqrt{\frac{t}{b}}\right)^m J_m(2\sqrt{tb}).
\end{eqnarray}

The integral can now be performed trivially as $\oint dz \, z^{k+m}=- 2 \pi \ii \,\delta_{k+m,-1}$, and $E(t)$ reads

\begin{eqnarray}
	&&E_F(t)= \mathcal{E}_0 e^{-\ii\omega_0 t}e^{-\Gamma't/2}\sum_{k=0}^\infty \frac{1}{(\Delta+\ii\Gamma'/2)^{k+1}}\nonumber\\ 
	&&\times\left(-\ii\sqrt{\frac{t}{b}}\right)^{-(k+1)} J_{-(k+1)}(2\sqrt{tb}).
\end{eqnarray}

Finally, given that $J_{-n}(x)=(-1)^nJ_n(x)$ and reintroducing $t_f$ back into the equation (the time at which the input pulse is switched off) by taking $t\rightarrow t-t_f$, we find

\begin{eqnarray}
\label{eq:bessel_sup1}
	&&\frac{I_\text{cont}}{I_0}=e^{-\Gamma' (t-t_f)}\big|\sum_{n=1}^\infty \left(\frac{-\ii}{\Delta+\ii\Gamma'/2}\sqrt{\frac{b}{(t-t_f)}}\right)^n\nonumber\\ 
	&&\times J_n\left(2\sqrt{ b (t-t_f)}\right)\big|^2,
\end{eqnarray}

where $I_0=|\mathcal{E}_0|^2$. Similarly, the transients at the beginning of the pulse are

 \begin{eqnarray}
\label{eq:bessel_sup2}
	&&\frac{I_\text{cont}}{I_0}=\big|t_\text{cont}(\omega_p)e^{-\ii \Delta (t-t_i)}-e^{-\Gamma' (t-t_i)/2}\nonumber\\ 
	&&\times\sum_{n=1}^\infty \left(\frac{-\ii}{\Delta+\ii\Gamma'/2}\sqrt{\frac{b}{(t-t_i)}}\right)^nJ_n\left(2\sqrt{ b (t-t_i)}\right)\big|^2.
\end{eqnarray}

The Bessel expansion in Eqs. \ref{eq:bessel_sup1} and \ref{eq:bessel_sup2} has a slow convergence for high optical depths. Alternatives to this expression are found in \cite{Macke13}.

\end{document}